\documentclass[english,showpacs,floatfix]{revtex4}
\usepackage[T1]{fontenc}
\usepackage[latin1]{inputenc}
\usepackage{amsmath}
\usepackage{amssymb}

\makeatletter
\@ifundefined{textcolor}{}
{%
 \definecolor{BLACK}{gray}{0}
 \definecolor{WHITE}{gray}{1}
 \definecolor{RED}{rgb}{1,0,0}
 \definecolor{GREEN}{rgb}{0,1,0}
 \definecolor{BLUE}{rgb}{0,0,1}
 \definecolor{CYAN}{cmyk}{1,0,0,0}
 \definecolor{MAGENTA}{cmyk}{0,1,0,0}
 \definecolor{YELLOW}{cmyk}{0,0,1,0}
}

\usepackage{longtable}\usepackage{dcolumn}\usepackage{babel}

\makeatother

\usepackage{babel}
\begin{document}

\title{Note on regular black holes in a brane world}

\author{J. C. S. Neves}

\email{nevesjcs@ime.unicamp.br}

\selectlanguage{english}%

\affiliation{Instituto de Matemática, Estatística e Computação Científica, Universidade
Estadual de Campinas \\
 CEP. 13083-859, Campinas-SP, Brazil}
\begin{abstract}
In this work, we show that regular black holes in a Randall-Sundrum-type
brane world model are generated by the nonlocal bulk influence, expressed
by a constant parameter in the brane metric, only in the spherical
case. In the axial case (black holes with rotation), this influence
forbids them. A nonconstant  bulk influence is necessary to generate regular black holes with rotation in this context.  
\end{abstract}

\pacs{04.70.Bw,04.20.Dw,04.20.Jb}

\maketitle

\section{Introduction}

The problem of singularities is an open issue in theoretical physics
[see, for example, Refs. \cite{Joshi,Novello} for discussions on this subject
in black hole (BH) physics and cosmology]. The general belief is that
only a complete quantum theory of gravity shall solve this issue.
Without a fully developed candidate for a quantum theory of gravity,
the singularities are avoided, for example, with some violations in
the energy conditions. These violations are more acceptable since
the discovery of cosmic accelerated expansion \cite{Supernova}. With
these violations, the Hawking-Penrose theorems are not valid. In cosmology or BH physics,
one may build solutions of the gravitational field equations without
a singularity. Such solutions may violate the Hawking-Penrose theorems
and avoid the problem of singularities.

The first regular black hole (RBH)\textemdash{}a compact object with
an event horizon and without a physical singularity\textemdash{}was
constructed by Bardeen in 1968 \cite{Bardeen2} in the general relativity
(GR) context as a development of ideas from Sakharov and co-workers
\cite{Sakharov} that spacetime in the black hole central region is
de Sitter-like. The Bardeen solution is spherically symmetric and
does not violate the weak energy condition (WEC). According to Ref. \cite{Beato},
it is formed by a nonlinear electrodynamic field. RBH solutions with
axial symmetry have been constructed as well (see the recent works  \cite{Smailagic,Modesto,Bambi,Neves_Saa,Toshmatov,Azreg,Larranaga2}).
Contrary to the spherical case, we have shown that axisymmetric RBHs always violate the WEC \cite{Neves_Saa}.

In this note, we study RBHs in a Randall-Sundrum-type model (RS). The
RS-I \cite{Randall-Sundrum}, two four-dimensional branes embedded
in a five-dimensional asymptotically anti-de Sitter (AdS) bulk, was introduced to solve the hierarchy problem in particle physics. The RS-II
\cite{Randall-Sundrum 2} adopts only one brane and was used in Ref. \cite{Shiromizu}
to obtain the induced gravitational field equations on the brane.
With these field equations, several authors \cite{Casadio,Bronnikov,Molina_Neves,Neves_Molina,Dadhich }
have studied BH solutions in this context. Among these solutions,
RBH solutions are generated. With spherical symmetry, previous works
have shown that RBHs may be obtained with a vacuum brane. The five-dimensional
bulk and its geometrical influence (a nonlocal influence) on the
brane provide a large class of metrics, including RBHs. On the other
hand, one shows that a nonlocal bulk effect, expressed by a constant
parameter in the brane metric, is incapable of generating RBHs with
a vacuum brane and forbids RBHs in the axial case. 

The structure of this paper is as follows: in Section
II, we present the gravitational field equations on the brane deduced
by Shiromizu, Maeda and Sasaki \cite{Shiromizu}; in Section III, we
show a regular solution with spherical symmetry on the brane constructed by Casadio \textit{et al}. \cite{Casadio}; in Section
IV, the axial case is presented and some properties are investigated; in Section V, the final remarks. We
adopt the metric signature $diag(-+++)$ and $c=1$, where $c$ is
the speed of light in vacuum.

\section{an approach to build black holes on the brane}

In the seminal work \cite{Shiromizu}, Shiromizu, Maeda and Sasaki obtained the gravitational field equations on the brane in a Randall-Sundrum-II-type
model. Adopting a four-dimensional brane embedded in a five-dimensional
asymptotically anti-de Sitter bulk and the $\mathbb{Z}_{2}$ symmetry,
the field equations on the brane are given by
\begin{equation}
G_{\mu\nu}=-\Lambda g_{\mu\nu}+8\pi G_{N}T_{\mu\nu}+\kappa_{5}^{4}\pi_{\mu\nu}-E_{\mu\nu},\label{field_equations}
\end{equation}
 where $\Lambda$, $T_{\mu\nu}$ and $g_{\mu\nu}$ are the cosmological
constant, the energy-momentum tensor and the metric on the brane,
respectively. $\kappa_{5}$ is a constant related to the Newtonian
constant, $G_{N},$ and the brane tension, $\lambda$, by $G_{N}=\frac{\kappa_{5}^{4}\lambda}{48\pi}$.
The tensor $\pi_{\mu\nu}$ is related to the energy-momentum tensor
on the brane: 
\begin{equation}
\pi_{\mu\nu}=-\frac{1}{4}T_{\mu\alpha}T_{\nu}^{\ \alpha}+\frac{1}{12}TT_{\mu\nu}+\frac{1}{8}g_{\mu\nu}T_{\alpha\beta}T^{\alpha\beta}-\frac{1}{24}g_{\mu\nu}T^{2}.\label{pi}
\end{equation}
 The traceless tensor ($E_{\mu}^{\mu}=0$ by construction) is the
electrical part of the five-dimensional Weyl tensor projected on the
brane. Following Ref. \cite{Maartens }, the tensor $E_{\mu\nu}$ gives
the nonlocal bulk influence on the four-dimensional spacetime. This
effect or influence on the brane may be described by the bulk free
gravitational radiation. The local influence is given by $\pi_{\mu\nu}$
on the matter fields. 

From Eq. (\ref{field_equations}), as $G_{\mu\nu}$ is the Einstein
tensor on the brane, one has the four-dimensional Ricci scalar
\begin{equation}
R=4\Lambda-8\pi G_{N}T_{\mu}^{\mu}-\kappa_{5}^{4}\pi_{\mu}^{\mu}.\label{R}
\end{equation}
Using a determined ansatz in Eq. (\ref{R}), with spherical or axial symmetry,
several authors have constructed BH solutions in this brane context.
In this paper we focus on regular solutions. In the next two sections
we present both spherical and axial solutions.

\section{the spherical case}

In the spherical case, one may solve Eq. (\ref{R}) using the general
spherical ansatz in the $(t,r,\theta,\phi)$ coordinates:
\begin{equation}
ds^{2}=-A(r)dt^{2}+\frac{1}{B(r)}dr^{2}+r^{2}(d\theta^{2}+sin^{2}\theta d\phi^{2}).\label{spherical_metric}
\end{equation}
The left side of (\ref{R}), with the aid of Eq. (\ref{spherical_metric}),
reads
\begin{equation}
R=2(1-B)-r^{2}B\left\{ \frac{A''}{A}-\frac{(A')^{2}}{2A^{2}}+\frac{A'B'}{2AB}+\frac{2}{r}\left[\frac{A'}{A}+\frac{B'}{B}\right]\right\} \,\,.\label{constraint}
\end{equation}
The symbol (') represents an ordinary derivative with respect to $r$.
Typically, Eq. (\ref{constraint}) was solved by some authors
by fixing a vacuum brane and it led to different geometries. $R=0$ \cite{Casadio,Bronnikov}
and $R=4\Lambda$ (with $\Lambda<0)$ \cite{Molina_Neves} led to
the RBHs. For example, in Ref. \cite{Casadio},
from the Schwarzschild metric element $A(r)=1-2m/r$, one has the
following solution of Eq. (\ref{constraint}):
\begin{equation}
B(r)=\left(1-\frac{2m}{r}\right)\left(\frac{1-\frac{3m}{2r}\left(1+\bar{\eta}\right)}{1-\frac{3m}{2r}}\right).\label{B_Casadio}
\end{equation}
With $A(r)$ and $B(r)$, one has a determined or fixed metric. In
the function (\ref{B_Casadio}), the constant $\bar{\eta}$ is related
to the post-Newtonian parameter $\beta$. Casadio \textit{et al.}
showed that the metric written using the $A(r)$ from Schwarzschild
geometry and $B(r)$ given by Eq.  (\ref{B_Casadio}) is regular everywhere
since $\bar{\eta}>0$. The Kretschmann invariant, for example, is
given by
\begin{equation}
K\sim\frac{\bar{\eta}^{2}}{\left(r-r_{s}\right)^{4}}.\label{K_Casadio}
\end{equation}
When $\bar{\eta}>0$ the coordinate system ends at $r=r_{0}>r_{s}$,
and the scalars or invariants are regular. In this example and others,
a vacuum brane with spherical symmetry can produce an RBH. This is
possible because of the nonlocal influence of the bulk on the brane.

\section{the axial case}

In the axial case, the Kerr-Schild ansatz is useful to solve Eq. (\ref{R}).
The general ansatz, with a cosmological constant, is given by
\begin{equation}
ds^{2}=ds_{\Lambda}^{2}+H\left(l_{\mu}dx^{\mu}\right)^{2},\label{Kerr-Schild}
\end{equation}
where $ds_{\Lambda}^{2}$ is the Minkowski $(\Lambda=0)$, pure anti-de
Sitter $(\Lambda<0)$ or de Sitter $(\Lambda>0)$ metric, $H$ is
a smooth function, and $l_{\mu}$ stands for a null vector. Its explicit
form in the $(\tau,r,\theta,\varphi)$ coordinates reads
\begin{equation}
ds_{\Lambda}^{2}=-\frac{(1-\frac{\Lambda}{3}r^{2})\Delta_{\theta}}{\Xi}d\tau^{2}+\frac{\Sigma}{(1-\frac{\Lambda}{3}r^{2})(r^{2}+a^{2})}dr^{2}+\frac{\Sigma}{\Delta_{\theta}}d\theta^{2}+\frac{(r^{2}+a^{2})sin^{2}\theta}{\Xi}d\varphi^{2}\label{ds_linha}
\end{equation}
 and
\begin{equation}
H(l_{\mu}dx^{\mu})^{2}=H\left(\frac{\Delta_{\theta}}{\Xi}d\tau+\frac{\Sigma}{(1-\frac{\Lambda}{3}r^{2})(r^{2}+a^{2})}dr-\frac{asin^{2}\theta}{\Xi}d\varphi\right)^{2},\label{H-1}
\end{equation}
 with
\begin{equation}
\Delta_{\theta}=1+\frac{\Lambda}{3}a^{2}cos^{2}\theta,\ \ \ \ \Sigma=r^{2}+a^{2}cos^{2}\theta,\ \ \ \ \Xi=1+\frac{\Lambda}{3}a^{2}.\label{definitions}
\end{equation}
The parameter $a$ plays the role of a rotational parameter. Substituting
Eqs. (\ref{ds_linha}) and (\ref{H-1}) into Eq. (\ref{R}), the left side
of Eq. (\ref{R}) is given by 
\begin{equation}
R=H''+\frac{4r}{\Sigma}H'+\frac{2}{\Sigma}H+4\Lambda.\label{R2}
\end{equation}
 Comparing Eq. (\ref{R2}) with Eq. (\ref{R}), one has
\begin{equation}
H''+\frac{4r}{\Sigma}H'+\frac{2}{\Sigma}H=-8\pi G_{N}T_{\mu}^{\mu}-\kappa_{5}^{4}\pi_{\mu}^{\mu}.\label{differential_eq}
\end{equation}
According to Refs. \cite{Aliev,Neves_Molina}, a vacuum brane $(T_{\mu\nu}=0)$
leads to the solution of the homogeneous version of Eq. (\ref{differential_eq}):
\begin{equation}
H=\frac{2mr}{\Sigma}-\frac{e}{\Sigma},\label{H2}
\end{equation}
where $m$ and $e$ are constants of integration. The parameter $m$
may be interpreted as the mass of the hole, and $e$ may be considered a tidal charge
or the geometrical nonlocal influence of the bulk on the brane (the
explicit form of $E_{\mu\nu}$ depends on $e$). However, the asymptotically flat, dS or AdS metrics with $H$ given by Eq. (\ref{H2})
are singular. Contrary to the spherical case, a constant mass function
does not lead to a regular metric in the axial case. With $m$ constant,
$R=4\Lambda$ and the Kretschmann scalar reads
\begin{equation}
K\sim\frac{m^{2}r^{6}}{\Sigma^{6}}+\frac{e^{2}r^{4}}{\Sigma^{6}},\label{K1}
\end{equation}
and one has a Kerr-like ring singularity inside the black hole (a
singular ring in the equatorial plane, $\theta=\pi/2$). To try generating
axisymmetric RBHs in a brane world context, one adopts the function
$H,$ used in Ref. \cite{Neves_Saa} to build RBHs in the GR context, plus
(or minus) the term which contains the constant tidal charge:
\begin{equation}
H=\frac{2m(r)r}{\Sigma}-\frac{e}{\Sigma}.\label{H3}
\end{equation}
With $m$ in Eq. (\ref{H2}) replaced by a general mass function $m(r)$,
the differential equation (\ref{differential_eq}) is not necessarily
homogeneous:
\begin{equation}
\frac{2\left(rm''(r)+2m'(r)\right)}{\Sigma}=-8\pi G_{N}T_{\mu}^{\mu}-\kappa_{5}^{4}\pi_{\mu}^{\mu}\neq0.\label{R3}
\end{equation}
That is, the brane is not a vacuum four-dimensional spacetime. The
quantity on the left in the above equation does not depend on $E_{\mu\nu}$
(because this tensor is traceless) and describes some matter field
on the brane. Then, a regular Ricci scalar is obtained from a mass
function such that for small $r$ the numerator of Eq. (\ref{R3}) must
vanish as $r^{\alpha}$ with $\alpha\geq2$ because $\Sigma$ depends
on $r^{2},$ according to Eq. (\ref{definitions}). With this condition,
the Ricci scalar is regular everywhere. [Note that the Eq. (\ref{R3})
is not in the GR context because $\pi_{\mu\nu}$ gives the local bulk
influence on the brane. In GR, the middle of Eq. (\ref{R3}) reads $-8\pi G_{N}T_{\mu}^{\mu}$]
In the same way, the Kretschmann scalar is given by
\begin{equation}
K\sim\frac{m(r)^{2}r^{6}}{\Sigma^{6}}+\frac{e^{2}r^{4}}{\Sigma^{6}}.\label{K2}
\end{equation}
However, a mass function of the type used in Ref. \cite{Neves_Saa} [for
small $r$, $m(r)\sim M_{0}r^{3},$ where $M_{0}$ is a constant]
does not prevent the existence of singularities in this brane world
context because the second term in Eq. (\ref{K2}) will diverge when $r\rightarrow0$
in the equatorial plane. When $e\neq0$, a mass function used in the
GR context does not work in this brane context because of the nonlocal
bulk influence, expressed by a constant parameter in the brane metric.
To obtain a regular metric on the brane, the parameter $e$ must be
replaced by a function $e(r)$, which for small $r$ behaves like
$e(r)\sim r^{4}$. 

\subsection{The spacetime structure}

To study the spacetime structure (the event and Killing horizons and ergosphere, for example) of a regular solution with $e(r)$ in Eq. (\ref{H3}), it is convenient to present the metric (\ref{Kerr-Schild}) in the Boyer-Lindquist coordinates ($t,r,\chi=cos\theta,\phi$). Making the coordinate transformations
\begin{eqnarray}
d\tau&=&dt+\frac{\Sigma H}{(1-\frac{\Lambda}{3}r^{2})\Delta_{r}}dr,\label{transf_1} \\
d\varphi&=&d\phi-\frac{\Lambda}{3}adt+\frac{a\Sigma H}{(r^{2}+a^{2})\Delta_{r}}dr,\label{transf_2}
\end{eqnarray}
 where  
\begin{equation}
\Delta_{r}=(r^{2}+a^{2})\left(1-\frac{\Lambda}{3}r^{2}\right)-2rm(r)+e(r),\label{Delta_r}
\end{equation}
the metric assumes the famous form:
\begin{eqnarray}
ds^{2} & = & -\frac{1}{\Sigma}\left(\Delta_{r}-\Delta_{\chi}a^{2}(1-\chi^2)\right)dt^{2}-\frac{2a}{\Xi\Sigma}\left[(r^{2}+a^{2})\Delta_{\chi}-\Delta_{r}\right](1-\chi^2) dtd\phi \nonumber \\
 & + & \frac{\Sigma}{\Delta_{r}}dr^{2}+\frac{\Sigma}{(1-\chi^2)\Delta_{\chi}}d\chi^{2}+\frac{1}{\Xi^{2}\Sigma}\left[(r^{2}+a^{2})^{2}\Delta_{\chi}-\Delta_{r}a^{2}(1-\chi^2)\right](1-\chi^2) d\phi^{2},\label{Metrica_Boyer-Lindquist}
\end{eqnarray}
with $\Delta_{\chi}=1+\frac{\Lambda}{3}a^2\chi^2$. 

Inspired by Ref. \cite{Neves_Saa}, which used a general mass which behaves like $m(r)\sim M_{0}r^3$ for small $r$ and $\lim_{r\rightarrow \infty}m(r)=M_0$, we have chosen a function $e(r)$ with the properties to avoid the central singularity, according to the above results, for small $r$, and an almost constant behaviour for large $r$, i.e.,  
\begin{equation}
e(r)=\frac{e_0}{(1+(\frac{r_0}{r})^q)^{\frac{4}{q}}}. \label{e}
\end{equation}  
The latter assumption is important to provide a metric close to the standard cases (Kerr-Newman-(A)-dS, for example) for large values of the radial coordinate. The constant $r_0$ is a length parameter, $e_0$ is a constant charge, and $q$ is a positive integer, the same used in the mass function. Due to the form of $e(r)$ for large values of the radial coordinate, this function does not play an important role in either the localization of the event horizon or the localization of the ergosphere. The horizons (inner and event) are obtained from roots of $\Delta_r (r_{\pm})=0$ or $g^{rr}(r_{\pm})=0$. The metric (\ref{Metrica_Boyer-Lindquist}) is independent of $t$ and $\phi$. Thus, it is endowed with the Killing vector fields $\xi_{t}=\frac{\partial}{\partial t}$ and $\xi_{\phi}=\frac{\partial}{\partial \phi}$. A Killing surface has null tangent Killing vectors and it is obtained from zeros of  $g_{tt}(S_{\pm})=0$. The expressions for both horizons and Killing surfaces are rather cumbersome. But the qualitative results are the same as those of Ref. \cite{Neves_Saa} in the GR context using only the general mass $m(r)$. That is, the nonconstant  bulk influence, $e(r)$, does not change the spacetime structure or the existence of horizons and Killing surfaces. In an illustrative form, we may indicate the spacetime structure at the equator ($\chi=0$), for the anti-de Sitter case, $\Lambda<0$, as
\begin{equation}
0<S_{-}<r_{-}<r_{+}<S_{+}<\infty.\label{AdS_region}
\end{equation}
And for the de Sitter case, $\Lambda >0$, the spacetime structure is given by  
\begin{equation}
0<S_{-}<r_{-}<r_{+}<S_{i}<S_{+}<r_{c}.\label{dS_region}
\end{equation}
The symbol $r_c$ represents the cosmological horizon, a present feature in asymptotically de Sitter geometries. The surface $S_i$ indicates the intermediate Killing horizon, which is another difference from the anti-de Sitter case. In the anti-de Sitter case, the ergosphere is localized between $S_+$ and $r_+$. In the de Sitter case, this important region is localized between $S_+$ and $S_i$. Both ergoregions have the same features of the Kerr ergosphere, i.e., the Killing vector $\xi_{t}$ is spacelike inside these regions.

\subsection{The weak energy condition}
In the GR context, as we said, RBHs with spherical symmetry, such as the Bardeen and Hayward geometries, do not violate the WEC. But, recently \cite{Neves}, using the approach of deforming black holes, we showed that RBHs with spherical symmetry may violate the WEC. On the other hand, when the rotation is present, according to Ref. \cite{Neves_Saa}, where the results were obtained in the GR context, this is not the case, and the WEC is always violated. 

To study the WEC violation in the axial case it is appropriate to write the energy-momentum tensor in a special frame, the locally nonrotating frame, used in Ref. \cite{Teukolsky} to investigate the Kerr black hole and its properties (in this frame, one has $T^{(a)(b)}=e^{(a)}_{\mu} e^{(b)}_{\nu}T^{\mu\nu}$). This was the approach adopted in Ref. \cite{Neves_Saa}. But in the brane world described by Eq. (\ref{field_equations}), the right side of the field equations gives the cosmological constant term plus the terms that contain the four-dimensional energy-momentum tensor and the tensor $E_{\mu\nu}$. The latter is not fully determined. To fix this tensor, the knowledge of the solution on the bulk is necessary  as well. Then, to try overcome this situation, we may interpret all terms on the right side of  Eq. (\ref{field_equations}) as an effective energy-momentum tensor in a four-dimensional interpretation ($T_{\mu\nu}^{eff}=-\Lambda g_{\mu\nu}+8\pi G_{N}T_{\mu\nu}+\kappa_{5}^{4}\pi_{\mu\nu}-E_{\mu\nu}$). But, in this case, we are not in a brane world context anymore.

\section{final remarks}

Regular black holes (RBHs) with spherical symmetry are obtained in
a Randall-Sundrum-type brane model in spite of the vacuum on the brane.
In this case, the nonlocal bulk influence on the brane\textemdash{}the geometrical influence expressed by a constant parameter in the brane metric\textemdash{}does not forbid its existence. On the other
hand, vacuum RBHs with axial symmetry (black holes with rotation)
are forbidden in this context. Moreover, the nonlocal bulk influence
must be expressed by a nonconstant  parameter in the brane metric
to produce RBHs with this symmetry in this brane world. This nonconstant  parameter must behave as $e(r)\sim r^4$ for small values of the radial coordinate. With this requirement, the Kretschmann scalar, for example, is finite everywhere.   

\begin{acknowledgments}
This work was supported by Fundação de Amparo à Pesquisa do Estado
de São Paulo (FAPESP), Brazil (Grant 2013/03798-3). I would like to
thank Alberto Saa and an anonymous referee for comments and suggestions.
 \end{acknowledgments}

\end{document}